\documentclass[twocolumn,showpacs,preprintnumbers,amsmath,amssymb,prl]{revtex4}

\usepackage{graphicx}
\usepackage{dcolumn}
\usepackage{bm}

\begin{document}

\preprint{APS/123-QED}

\title{A Search for Variations of Fundamental Constants using Atomic Fountain Clocks}
\author{H. Marion}
\author{F. Pereira Dos Santos}
\author{M. Abgrall}
\author{S. Zhang}
\author{Y. Sortais}
\author{S. Bize}
\author{I. Maksimovic}
\author{D. Calonico}
    \altaffiliation[Present address: ]{Istituto Elettrotecnico Nazionale "G. Ferraris" , Strada delle Cacce 41, 10135 Torino, Italy}
\author{J. Gr\"unert}
\author{C. Mandache}
    \altaffiliation[Present address: ]{Institutul National de Fizica Laserilor, Plasmei si Radiatiei, P.O. Box MG36, Bucaresti, Magurele, Romania}
\author{P. Lemonde}
\author{G. Santarelli}
\author{Ph. Laurent}
\author{A. Clairon}
    \affiliation{BNM-SYRTE, Observatoire de Paris, 61 Avenue de l'Observatoire, 75014 Paris, France }
\author{C. Salomon}
    \affiliation{Laboratoire Kastler Brossel, ENS, 24 rue Lhomond, 75005 Paris, France}

\date{\today}

\begin{abstract}Over five years we have compared the hyperfine frequencies of
$^{133}\text{Cs}$ and $^{87}\text{Rb}$ atoms in their electronic
ground state using several laser cooled $^{133}\text{Cs}$ and
$^{87}\text{Rb}$ atomic fountains with an accuracy of $\sim
10^{-15}$. These measurements set a stringent upper bound to a
possible fractional time variation of the ratio between the two
frequencies :
$\frac{d}{dt}\ln\left(\frac{\nu_{\text{Rb}}}{\nu_{\text{Cs}}}\right)=(0.2\pm
7.0)\times 10^{-16}\,\text{yr}^{-1}$ ($1\sigma$ uncertainty). The
same limit applies to a possible variation of the quantity
$(\mu_{\text{Rb}}/\mu_{\text{Cs}})\alpha^{-0.44}$, which involves
the ratio of nuclear magnetic moments and the fine structure
constant.
\end{abstract}


\maketitle

Since Dirac's 1937 formulation of his large number hypothesis
aiming at tying together the fundamental constants of physics
\cite{Dirac37}, large amount of work has been devoted to test if
these constants were indeed constant over time
\cite{Dyson,Uzan02}.

In General Relativity and in all metric theories of gravitation,
variations with time and space of non gravitational fundamental
constants such as the fine structure constant
$\alpha=e^2/4\pi\epsilon_0\hbar c$ are forbidden. They would
violate Einstein's Equivalence Principle (EEP). EEP imposes the
Local Position Invariance stating that in a local freely falling
reference frame, the result of any local non gravitational
experiment is independent of where and when it is performed. On
the other hand, almost all modern theories aiming at unifying
gravitation with the three other fundamental interactions predict
violation of EEP at levels which are within reach of near-future
experiments \cite{Damour94,Calmet02}. As the internal energies of
atoms or molecules depend on electromagnetic, as well as strong
and weak interactions, comparing the frequency of electronic
transitions, fine structure transitions and hyperfine transitions
as a function of time or gravitational potential provides an
interesting test of the validity of EEP.

To date, very stringent tests exist on geological and cosmological
timescales. The analysis of the Oklo nuclear reactor showed that,
$2\times 10^9$\, years ago, $\alpha$ did not differ from the
present value by more than $10^{-7}$ of its value \cite{Damour96}.
Light emitted by distant quasars has been used to perform
absorption spectroscopy of interstellar clouds. For instance,
measurements of the wavelengths of molecular hydrogen transitions
test a possible variation of the electron to proton mass ratio
$m_e/m_p$ \cite{Ivanchik02}. Comparisons between the gross
structure and the fine structure of neutral atoms and ions would
indicate that $\alpha$ for a redshift $z\sim 1.5$ ($\sim 10$ Gyr)
differed from the present value: $\Delta \alpha/\alpha=(-7.2\pm
1.8)\times 10^{-6}$ \cite{Webb01}. Today this is the only claim
that fundamental constants might change.

On much shorter timescales, several tests using frequency
standards have been performed
\cite{Turneaure83,Godone93,Prestage95}. These laboratory tests
have a very high sensitivity to changes in fundamental constants.
They are repeatable, systematic errors can be tracked as
experimental conditions can be changed.

In this letter we present results that place a new stringent limit
to the time variation of fundamental constants. By comparing the
hyperfine energies of $^{133}$Cs and $^{87}$Rb in their electronic
ground state over a period of nearly five years, we place an upper
limit to the rate of change of the ratio of the hyperfine
frequencies $\nu_{\text{Rb}}/\nu_{\text{Cs}}$. Our measurements
take advantage of the high accuracy ($\sim 10^{-15}$) of several
laser cooled Cs and Rb atomic fountains. According to recent
atomic structure calculations \cite{Prestage95,Dzuba99}, these
measurements are sensitive to a possible variation of the quantity
$(\mu_{\text{Rb}}/\mu_{\text{Cs}})\alpha^{-0.44}$, where $\mu$'s
are the nuclear magnetic moments. We anticipate major advances in
these tests using frequency standards, thanks to recent advances
in optical frequency metrology using femtosecond lasers
\cite{StAndrews,Bize02}.

\begin{figure}[htb]
\includegraphics[height=3.5cm]{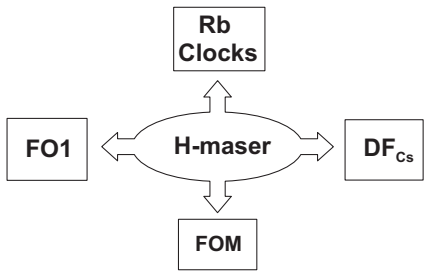}
\caption{BNM-SYRTE clock ensemble.  A single 100\,MHz signal from
a H-maser is used for frequency comparisons and is distributed to
each of the microwave synthesizers of the $^{133}$Cs (FO1, FOM,
DF$_{\text{Cs}}$) and $^{87}$Rb fountain clocks. In 2001, the Rb
fountain has been upgraded and is now a dual fountain using
alternately rubidium (DF$_{\text{Rb}}$) or cesium atoms
(DF$_{\text{Cs}}$). } \label{fig:clocks}
\end{figure}

In our experiments, three atomic fountains are compared to each
other, using a hydrogen maser (H-maser) as a flywheel oscillator
(Fig.\ref{fig:clocks}). Two fountains, a transportable fountain
FOM, and FO1 \cite{FO1} are using cesium atoms. The third fountain
is a dual fountain (DF) \cite{Bize01}, operating alternately with
rubidium (DF$_{\text{Rb}}$) and cesium (DF$_{\text{Cs}}$). These
fountains have been continuously upgraded in order to improve
their accuracy from $2\times 10^{-15}$ in 1998 to $8\times
10^{-16}$ for cesium and from $1.3 \times 10^{-14}$ \cite{Bize99}
to $6\times 10^{-16}$ for rubidium.

Fountain clocks operate as follows. First, atoms are collected and
laser cooled in an optical molasses or in a magneto-optical trap
in $0.3$ to $0.6$ s. Atoms are launched upwards, and selected in
the clock level ($m_F=0$) by a combination of microwave and laser
pulses. Then, atoms interact twice with a microwave field tuned
near the hyperfine frequency, in a Ramsey interrogation scheme.
The microwave field is synthesized from a low phase noise 100 MHz
signal from a quartz oscillator, which is phase locked to the
reference signal of the H-maser (Fig.\ref{fig:clocks}). After the
microwave interactions, the population of each hyperfine state is
measured using light induced fluorescence. This provides a
measurement of the transition probability as a function of
microwave detuning. Successive measurements are used to steer the
average microwave field to the frequency of the atomic resonance
using a digital servo system. The output of the servo provides a
direct measurement of the frequency difference between the H-maser
and the fountain clock.

The three fountains have different geometries and operating
conditions: the number of detected atoms ranges from $3\times
10^{5}$ to $2\times 10^{6}$ at a temperature of $\sim 1\,\mu$K,
the fountain cycle duration from 1.1 to 1.6 s. The Ramsey
resonance width is between 0.9 and $1.2$ Hz. In measurements
reported here the fractional frequency instability is
$(1-2)\times10^{-13}\tau^{-1/2}$, where $\tau$ is the averaging
time in seconds. Fountain comparisons have a typical resolution of
$\sim 10^{-15}$ for a 12 hour integration, and each of the four
data campaigns lasts from 1 to 2 months during which an accuracy
evaluation of each fountain is performed.

\begin{figure}[htb]
\begin{center}
\includegraphics[height=8cm]{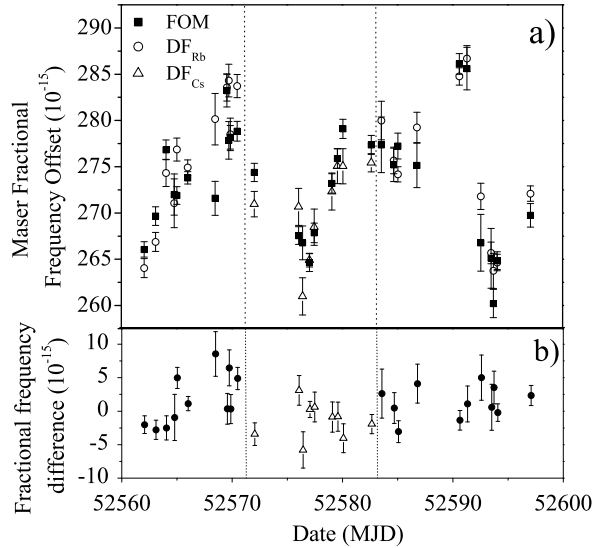}
\end{center}
\caption{The 2002 frequency comparison data. a) H-maser fractional
frequency offset versus FOM ($\blacksquare$), and alternately
versus DF$_{\text{Rb}}$ ($\circ$) and DF$_{\text{Cs}}$
($\vartriangle$ between dotted lines). b) Fractional frequency
differences. Between dotted lines, Cs-Cs comparisons, outside
Rb-Cs comparisons. Error bars are purely statistical. They
correspond to the Allan standard deviation of the comparisons and
do not include contributions from fluctuations of systematic
shifts of Table \ref{tab:Budget}.} \label{fig:compRbCs}
\end{figure}
The 2002 measurements are presented in Fig.\ref{fig:compRbCs},
which displays the maser fractional frequency offset, measured by
the Cs fountains FOM and DF$_{\text{Cs}}$. Also shown is the
H-maser frequency offset measured by the Rb fountain
DF$_{\text{Rb}}$ where the Rb hyperfine frequency is
conventionally chosen to be
$\nu_{\text{Rb}}(1999)=6\,834\,682\,610.904\,333\,$Hz, our 1999
value. The data are corrected for the systematic frequency shifts
listed in Table \ref{tab:Budget}. The H-maser frequency exhibits
fractional fluctuations on the order of $10^{-14}$ over a few
days, ten times larger than the typical statistical uncertainty
resulting from the instability of the fountain clocks. In order to
reject the H-maser frequency fluctuations, the fountain data are
recorded simultaneously (within a few minutes). The fractional
frequency differences plotted in Fig.\ref{fig:compRbCs}\,b
illustrate the efficiency of this rejection. DF is operated
alternately with Rb and Cs, allowing both Rb-Cs comparisons and
Cs-Cs comparisons (central part of Fig.\ref{fig:compRbCs}) to be
performed.

\begin{table}[htb]
\caption{Accuracy budget of the fountains involved in the 2002
measurements (DF et FOM).}
\begin{center}
\begin{tabular}{cccc}
\textbf{Fountain} & \multicolumn{1}{|c}{\textbf{DF$_{\text{Cs}}$}} & \multicolumn{1}{|c}{\textbf{DF$_{\text{Rb}}$}} & \multicolumn{1}{|c}{\textbf{FOM}}  \\
\hline\hline Effect & \multicolumn{3}{|c}{ Value \& Uncertainty  (10$^{-16})$} \\
\hline 2$^{nd}$\ order Zeeman &  \multicolumn{1}{|r} {$1773.0\pm5.2$} & \multicolumn{1}{|r} {$3207.0 \pm 4.7$} & \multicolumn{1}{|r} {$385.0 \pm 2.9$}\\
\hline Blackbody Radiation & \multicolumn{1}{|r} {$-173.0\pm 2.3$}& \multicolumn{1}{|r} {$-127.0\pm 2.1$}& \multicolumn{1}{|r} {$-186.0\pm 2.5$}\\
\hline {
\begin{tabular}{r}Cold collisions \\
+ cavity pulling
\end{tabular}
}
 & \multicolumn{1}{|r} {$-95.0\pm 4.6$}& \multicolumn{1}{|r} {$0.0\pm 1.0$}& \multicolumn{1}{|r} {$-24.0\pm 4.8$}\\
\hline others &  \multicolumn{1}{|r} {$0.0\pm 3.0$}& \multicolumn{1}{|r} {$0.0\pm 3.0$}& \multicolumn{1}{|r} {$0.0\pm 3.7$}\\
\hline \hline \textbf{Total uncertainty} & \multicolumn{1}{|r}
{\textbf 8} & \multicolumn{1}{|r} {\textbf 6} &
\multicolumn{1}{|r} {\textbf 8}
\end{tabular}
\end{center}
\label{tab:Budget}
\end{table}
Systematic effects shifting the frequency of the fountain
standards are listed in Table \ref{tab:Budget}. The quantization
magnetic field in the interrogation region is determined with a
$0.1$~nT uncertainty by measuring the frequency of a linear
field-dependent ``Zeeman" transition. The temperature in the
interrogation region is monitored with 5 platinum resistors and
the uncertainty on the black-body radiation frequency shift
corresponds to temperature fluctuations of about 1 K
\cite{Simon98}. Clock frequencies are corrected for the cold
collision and cavity pulling frequency shifts using several
methods \cite{Pereira02,Sortais00}. All other effects do not
contribute significantly and their uncertainties are added
quadratically. We searched for the influence of synchronous
perturbations by changing the timing sequence and the atom launch
height. To search for possible microwave leakage, we changed the
power ($\times9$) in the interrogation microwave cavity. No shift
was found at a resolution of $10^{-15}$. The shift due to residual
coherences and populations in neighboring Zeeman states is
estimated to be less than $10^{-16}$. As shown in \cite{Wolf01},
the shift due to the microwave photon recoil is very similar for
Cs and Rb and smaller than $+ 1.4\times 10^{-16}$. Relativistic
corrections (gravitational redshift and second order Doppler
effect) contribute to less than $10^{-16}$ in the clock
comparisons.

For the Cs-Cs 2002 comparison, we find:
\begin{equation}
\frac{{\nu_{\text{Cs}}^{\mathrm{DF}}(2002)}-\nu_{\text{Cs}}^{\mathrm{FOM}}(2002)}{\nu_{\text{Cs}}}=+12(6)(12)\times
10^{-16}
\end{equation}
where the first parenthesis reflects the $1\sigma$ statistical
uncertainty, and the second the systematic uncertainty, obtained
by adding quadratically the inaccuracies of the two Cs clocks (see
Table \ref{tab:Budget}). The two Cs fountains are in good
agreement despite their significantly different operating
conditions (see Table \ref{tab:Budget}), showing that systematic
effects are well understood at the $10^{-15}$ level.

In 2002, the $^{87}$Rb frequency measured with respect to the
average $^{133}$Cs frequency is found to be:
\begin{equation}
\nu_{\text{Rb}}(2002)=6\,834\,682\,610.904\,324(4)(7) \,\text{Hz}
\end{equation}
where the error bars now include DF$_{\text{Rb}}$,
DF$_{\text{Cs}}$ and FOM uncertainties. This is the most accurate
frequency measurement to date.

\begin{figure}[htb]
\begin{center}
\includegraphics[height=6cm]{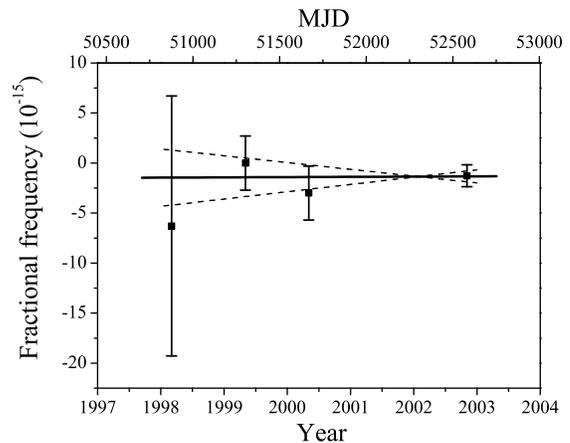}
\end{center}
\caption{Measured $^{87}$Rb frequencies referenced to the
$^{133}$Cs fountains over 57 months. The 1999 measurement value
($\nu_{\text{Rb}}(1999)=6\,834\,682\,610.904\,333\,$Hz) is
conventionally used as reference. A weighted linear fit to the
data gives
$\frac{d}{dt}\ln\left(\frac{\nu_{\text{Rb}}}{\nu_{\text{Cs}}}\right)=(0.2\pm
7.0)\times 10^{-16}\,\text{yr}^{-1}$. Dotted lines correspond to
the $1\sigma$ slope uncertainty.} \label{fig:alphapoint}
\end{figure}
In Fig.\ref{fig:alphapoint} are plotted all our Rb-Cs frequency
comparisons. Except for the less precise 1998 data \cite{Bize99},
two Cs fountains were used together to perform the Rb
measurements. The uncertainties for the 1999 and 2000 measurements
were $2.7\times 10^{-15}$, because of lower clock accuracy and
lack of rigorous simultaneity in the earlier frequency comparisons
\cite{Bize01}. A weighted linear fit to the data in
Fig.\ref{fig:alphapoint} determines how our measurements constrain
a possible time variation of $ \nu_{\text{Rb}}/\nu_{\text{Cs}}$.
We find:
\begin{equation} \label{eq:alpha1}
\frac{d}{dt}\ln\left(\frac{\nu_{\text{Rb}}}{\nu_{\text{Cs}}}\right)=(0.2\pm7.0)\times
10^{-16}\,\text{yr}^{-1}
\end{equation}
which represents a 5-fold improvement over our previous results
\cite{Bize01} and a 100-fold improvement over the Hg$^+$-H
hyperfine energy comparison \cite{Prestage95}.

We now examine how this result constrains possible variations of
fundamental constants. For an alkali with atom number $Z$, the
hyperfine transition frequency can be approximated by:
\begin{equation}
\nu \propto
\alpha^2\frac{\mu}{\mu_N}\left(\frac{m_e}{m_p}\right)R_\infty
c~F_{rel}(Z\alpha),
\end{equation}
where $R_\infty$ is the Rydberg constant, $c$ the speed of light,
$\mu$ the magnetic moment of the nucleus, $\mu_N$ the nuclear
magneton. $F_{rel}(Z\alpha)$ is a relativistic function which
strongly increases with $Z$ \cite{Prestage95,Karshenboim00}. For
$^{133}$Cs, this Casimir relativistic contribution amounts to
40\,\% of the hyperfine splitting and $\alpha \frac{\partial
ln(F_{rel}(Z\alpha))}{\partial\alpha}=0.74$. For $^{87}$Rb, this
quantity is 0.30 \cite{ref}. Following \cite{Prestage95} and
neglecting possible changes of the strong and weak interactions
affecting $\mu_{\text{Rb}}$ and $\mu_{\text{Cs}}$, the sensitivity
of the ratio $\nu_{\text{Rb}}/\nu_{\text{Cs}}$ to a variation of
$\alpha$ is simply given by:
\begin{equation}\label{eq:alpha2}
\frac{\partial}{\partial\ln\alpha
}\ln\left(\frac{\nu_{\text{Rb}}}{\nu_{\text{Cs}}}\right)\simeq
(0.30-0.74)=-0.44.
\end{equation}
Using equations \ref{eq:alpha1} and \ref{eq:alpha2}, we thus set
the new limit: $\dot{\alpha}/\alpha=(-0.4\pm16)\times
10^{-16}\,\text{yr}^{-1}$.

In contrast with \cite{Prestage95}, Ref.\cite{Karshenboim00}
argues that a time variation of the nuclear magnetic moments must
also be considered in a comparison between hyperfine frequencies.
The magnetic moments $\mu$ can be calculated using the Schmidt
model. For atoms with odd $A$ and $Z$ such as $^{87}$Rb and
$^{133}$Cs, the Schmidt magnetic moment $\mu^{(s)}$ is found to
depend only on $g_p$, the proton gyromagnetic ratio. With this
simple model, Ref.\cite{Karshenboim00} finds:
\begin{equation}\label{eq:gp}
\frac{\partial}{\partial\ln g_p
}\ln\left(\frac{\nu_{\text{Rb}}}{\nu_{\text{Cs}}}\right)\simeq
\frac{\partial}{\partial \ln
g_p}\ln\left(\frac{\mu_{\text{Rb}}^{(s)}}{\mu_{\text{Cs}}^{(s)}}\right)\simeq
2.0.
\end{equation}
Attributing any variation of $\nu_{\text{Rb}}/\nu_{\text{Cs}}$ to
a variation of $g_p$, equations \ref{eq:alpha1} and \ref{eq:gp}
lead to: $\dot{g_p}/g_p=(0.1\pm3.5)\times
10^{-16}\,\text{yr}^{-1}$. However, it must be noted that the
Schimdt model is over simplified and does not agree very
accurately with the actual magnetic moment.

Moreover, attributing all the time variation of
$\nu_{\text{Rb}}/\nu_{\text{Cs}}$ to either $g_p$ or $\alpha$
independently is somewhat artificial. Theoretical models allowing
for a variation of $\alpha$ also allow for variations in the
strength of the strong and electroweak interactions. For instance,
Ref.\cite{Calmet02} argues that Grand unification of the three
interactions implies that a time variation of $\alpha$ necessarily
comes with a time variation of the coupling constants of the other
interactions. Ref.\cite{Calmet02} predicts that a fractional
variation of $\alpha$ is accompanied with a $\sim 40$ times larger
fractional change of $m_e/m_p$. In order to independently test the
stability of the three fundamental interactions, several
comparisons between different atomic species and/or transitions
are required. For instance and as illustrated in \cite{Bize02},
absolute frequency measurements of an optical transition is
sensitive to a different combination of fundamental constants:
$(\mu_{\text{Cs}}/{\mu_N})(m_e/m_p)\alpha^x$, where $x$ depends on
the particular atom and/or transition.

A more complete theoretical analysis going beyond the Schmidt
model would clearly be very useful to interpret frequency
comparisons involving hyperfine transitions. This is especially
important as most precise frequency measurements, both in the
microwave and the optical domain \cite{Bize02,Niering00,Udem01},
are currently referenced to the $^{133}$Cs hyperfine splitting,
the basis of the SI definition of the second. The H hyperfine
splitting, which is calculable to a high accuracy, has already
been considered as a possible reference several decades ago.
Unfortunately, despite numerous efforts, the H hyperfine splitting
is currently measured to only 7 parts in $10^{13}$ (using
H-masers), almost three orders of magnitude worse than the results
presented in this letter.

In summary, by comparing $^{133}$Cs and $^{87}$Rb hyperfine
energies, we have set a stringent upper limit to a possible
fractional variation of the quantity
$(\mu_{\text{Rb}}/\mu_{\text{Cs}})\alpha^{-0.44}$ at $(-0.2\pm
7.0)\times 10^{-16}\text{yr}^{-1}$. In the near future, accuracies
near 1 part in $10^{16}$ should be achievable in microwave atomic
fountains, improving our present Rb-Cs comparison by one order of
magnitude.

We anticipate that comparisons between rapidly progressing optical
and microwave laser-cooled frequency standards currently developed
in several laboratories will bring orders of magnitude gain in
sensitivity. In order to have the full benefit of these advances,
frequency comparisons with improved accuracy between these distant
clocks will be necessary. Serving this purpose, a new generation
of time/frequency transfer at the $10^{-16}$ level is currently
under development for the ESA space mission ACES which will fly
ultra-stable clocks on board the international space station in
2006 \cite{Salomon01}. These comparisons will also allow for a
search of a possible change of fundamental constants induced by
the annual modulation of the Sun gravitational potential due to
the elliptical orbit of the Earth \cite{Bauch02}.

The authors wish to thank T. Damour, J.P. Uzan, and P. Wolf for
valuable discussions, A. G\'erard and the electronic staff for
technical assistance. This work was supported in part by BNM, CNRS
and CNES. BNM-SYRTE and Laboratoire Kastler Brossel are Unit\'es
Associ\'ees au CNRS, UMR 8630 and 8552.


\begin{thebibliography}{30}
\bibitem{ref} A more precise calculation in \cite{Dzuba99} gives $LdF_{rel}(^{133}\rm{Cs})=0.83$, which differs by 10\% from the Casimir formula.
\end{thebibliography}
\end{document}